\documentclass[12pt]{article}           

\def\title{A comment on causality}

\long\def\abstract{
We start from the well-known form of the interval of the special
relativity, stare it, and build up an attempt to implement the causality 
from it. Some features appear to be new, they involve the mass of 
the particle and the structure of space-time.}

\input{epsf}
\input{axodraw.sty}

\begin{document}

\vskip 21mm
\centerline{\huge\bf A comment on causality.}
\vskip 7mm
\begin{center} \vskip 14mm
Ricardo Bent\'{\i}n{\begingroup\def\thefootnote{*}
                                \footnote{e-mail: rbentin@ift.unesp.br}
                                \addtocounter{footnote}{-1} \endgroup}
 \vskip 21mm
\end{center}

\begin{center} {\bf ABSTRACT } \end{center}   
\abstract \vfill \noindent \\
\bf{\tiny erev Tishrei, 10th.}
\thispagestyle{empty} \newpage
\pagestyle{plain} 
\newpage
\setcounter{page}{1}

Causality supports the construction of several modern theories in 
physics \cite{siegel}.
Its interests abroad and overcome the very beginning of special relativity and
goes to quantum mechanics at S-matrix level when the unitarity condition
$$
   {\cal S}^\dag{\cal S} =1
$$
does not means causality, rather it just means time development which
conserves probability, if we want to incorporate a causal description
we have to deal with the chronological order, but this is also a weak
condition. 
Another manner of state causality independent of special relativity and 
quantum mechanics is via the concept of locality, i.e., all interactions
occur at a point so there is no action at a distance, and this is used
by relativistic quantum field theory.
As an example, in the context of quantum field theories, the choice of 
the light cone gauge leads in pathological non physical problems. The
origin of these problems lays in considerations of non causal distributions
and they disappear when causality is restored \cite{first}.
There is also defined the microscopic causality as the idea, based on 
special relativity, in which information cannot travel faster than the 
speed of light and this is used to demonstrate the spin 
statistics theorem.\\
We are not defying the other previous ideas of implement causality in 
a theory, we just wonder to have an approximation from another perspective 
to deem causality. Our analyze is as simple as possible, so we are not
going into quantum theory neither general relativity. Besides we are just
considering the theory of special relativity as our arena and we are 
going to make use of de Sitter spacetime, so in the first section we
will summarize the de Sitter and Anti de Sitter spacetimes. The next 
section is devoted to explain our form of implement causality.
\section{Review of (Anti)de Sitter spacetime.}
In this section we will make a briefly review of Anti de Sitter and
de Sitter spacetime in four dimensions, 
that belong to the set of constant curvature
spacetime metrics. This kind of metrics are locally characterized
by the condition 
$$
  R_{abcd}=\frac{1}{12}R(g_{ac}g_{bd}-g_{ad}g_{bc}),
$$
where the indexes runs over $0,..,3$. This equation is
equivalent to $C_{ab}=0=R_{ab}-\frac{1}{4}Rg_{ab}$, and this implies
that the Riemann tensor is completely determined by the Ricci 
scalar $R$. Using Bianchi identities we can conclude that $R$ is constant
over all spacetime. In this way we can also classified this spacetime
according the sign of $R$ as
\begin{list}{}{}
\item $R<0$ Anti de Sitter spacetime,
\item $R=0$ Minkowski spacetime,
\item $R>0$ de Sitter spacetime.
\end{list}

\subsection{De Sitter spacetime.}
Here we will brusquely describe the above case for $R>0$
i.e., the de Sitter spacetime, which has the $R^1xS^3$ topology.
It can be visualized as the hyperboloid
$$
  -v^2+w^2+x'^2+y'^2+z'^2=\alpha^2
$$
in a five dimensional space $R^5$ with metric
\begin{equation}
  \label{metric}
  -dv^2+dw^2+dx'^2+dy'^2+dz'^2=ds^2.
\end{equation}
Introducing the coordinates $(\hat{t},\xi,\theta,\phi)$ on the hyperboloid
with the relations
\begin{eqnarray*}
  \alpha sinh(t/\alpha)=v,\\
  \alpha cosh(t/\alpha)cos\xi=w,\\
  \alpha cosh(t/\alpha)sin\xi cos\theta=x',\\
  \alpha cosh(t/\alpha)sin\xi sin\theta cos\phi=y',\\
  \alpha cosh(t/\alpha)sin\xi sin\theta sin\phi=z',
\end{eqnarray*}
using these coordinates, the metric takes the following
form
$$
  ds^2=-d\hat{t}^2+\alpha^2cosh^2(t/\alpha)
    [d\xi^2+sin^2(\xi)(d\theta^2+sin^2\theta d\phi^2)],
$$
and this is called the spherical coordinates.
There is another interesting form of the metric named planar
coordinates in which
\begin{eqnarray*}
  t&=&\alpha\ln\frac{v+w}{\alpha},\\
  x&=&\alpha\frac{x'}{v+w},\\
  y&=&\alpha\frac{y'}{v+w},\\
  z&=&\alpha\frac{z'}{v+w},
\end{eqnarray*}
then, the metric is
$$
  -dt^2+e^{2t/\alpha}.(dx^2+dy^2+dz^2)\doteq d{\cal S}^2,
$$
We can also be able to see that sections of constant $t''$ are spheres 
$S^3$ with positive curvature, where we define the temporal 
coordinate $t''$ as
$$
  t''=2arctan(e^{t'/\alpha})-\frac{\pi}{2}, \ \ \
-\frac{\pi}{2}<t''<\frac{\pi}{2}.
$$
then the metric turns to be
$$
  ds^2=\alpha^2cosh^2(t'/\alpha)
  [-dt''^2+d\xi^2+sin^2\xi(d\theta^2+sin^2\theta d\phi^2)].
$$
This shows us that the de Sitter spacetime is conformal to the 
Einstein's static universe which is defined by the metric
$$
  ds^2=-dt''^2+dr'^2+sin^2r'(d\theta^2+sin^2\theta d\phi^2).
$$
This describes a universe of positive cosmological constant
$\Lambda > 0$.

\subsection{Anti de Sitter spacetime.}
The space of constant negative curvature, $R<0$, is denominated as
Anti de Sitter spacetime. It has the $S^1xR^3$ topology and can be
represented as being the hyperboloid
$$
  -u^2-v^2+x^2+y^2+z^2=1,
$$
over the flat five dimensional space $R^5$ with the associated 
metric
$$
  -du^2-dv^2+dx^2+dy^2+dz^2=ds^2.
$$
The Anti de Sitter spacetime describes a universe of negative
$\Lambda$ \cite{large}.

\section{\large Causality implementation.}
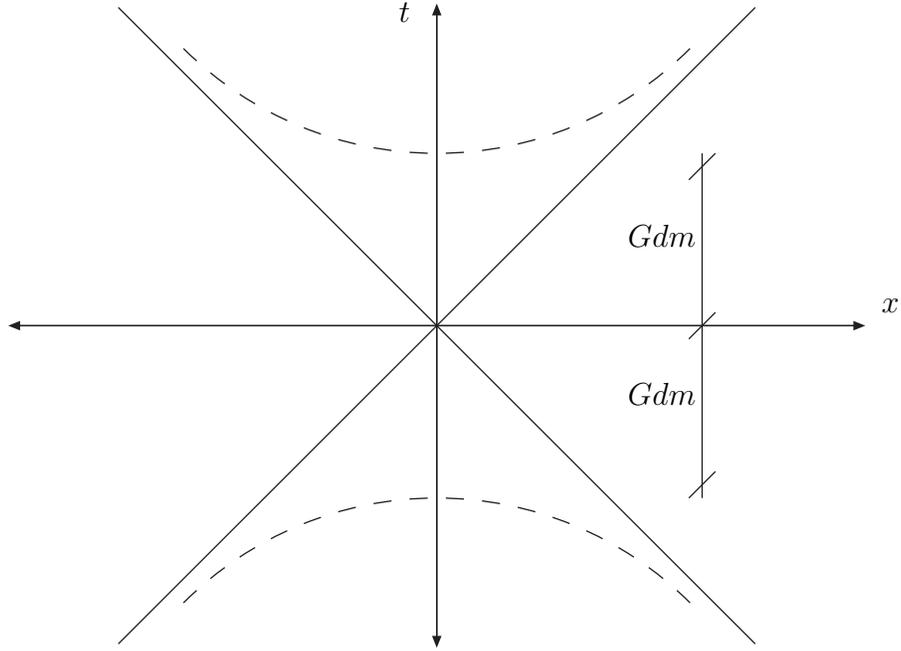
\begin{figure}[htb]
\centerline{
\begin{picture}(120,240)(-60,-120)
\put(175,5){\makebox(0,0)[br]{$x$}}
\put(-10,115){\makebox(0,0)[br]{$t$}}
\LongArrow(-160,0)(160,0)
\LongArrow(160,0)(-160,0)
\LongArrow(0,-120)(0,120)
\LongArrow(0,120)(0,-120)
\Line(120,120)(-120,-120)
\Line(120,-120)(-120,120)
\DashCArc(0,200)(135,225,315)7
\DashCArc(0,-200)(135,45,135)7
\Line(100,65)(100,-65)
\Line(95,55)(105,65)
\put(98,30){\makebox(0,0)[br]{$Gdm$}}
\Line(95,-5)(105,5)
\put(98,-30){\makebox(0,0)[br]{$Gdm$}}
\Line(95,-65)(105,-55)
\end{picture}
}
\caption{Light-cone and causal-hyperboloid.}
\label{lcch}
\end{figure}

There are many ways in which people implement causality, in 
this work we will not fret over them, instead we are going to
move the scenario on to special relativity where for every Lorentz
frame the event $B$ is in the future of event $A$ if and only if
$B$ is in $A$'s light cone.
Starting from the relativistic expression of the interval and using
a flat metric with signature $(-,+,+,+)$:
\begin{eqnarray}
        ds^2&=&-dt^2+dx^2+dy^2+dz^2.
\label{inter}   
\end{eqnarray}
Where we are working in units of $c=h=1$. Then the {\it
mathematical}
condition for a particle remains {\it inside} or {\it on} the light-cone 
is expressed as:
\begin{eqnarray}
        (ds)^2&\le&0
\label{ineq}
\end{eqnarray}
This means that a space-like or time-like particle can not have a
speed higher than $c$.
Inequation (\ref{ineq}) just says that. But it also has some taste of
causality. The problem is, perhaps, the fact that working with
an inequation does not give further information about it. The question is:
how to use (\ref{ineq}) in order to manipulate causality?. We will
propose an answer:\\
Since $(ds)^2$ is a quadratic expression and also $(ds)^2 \le 0$, then it
will be equal to another physical-meaning quadratic expression
multiplied by a negative sign, this guarantees the correct sign we 
want for inequation (\ref{ineq}) be satisfied.\\
Using (\ref{inter}) and (\ref{ineq}), it would read something like this:
\begin{eqnarray}
\nonumber       ds^2&=&-A^2\\
                        -dt^2+dx^2+dy^2+dz^2&=&-A^2,
\label{a}
\end{eqnarray}
where $A$ is a quantity to be determined from physics.\\
Now, another question arises now: what is the meaning of the quantity
$A$?. The answers is not so easy as before, but let's try to
figure out something.\\
We should classified our world as something where we have {\it massive} and
other {\it massless} objects like the protons and photons respectively.\\
From Einstein's observations, no mass can live {\it on} the light-cone,
it must lie {\it inside} it!, i.e., the inequation (\ref{ineq}) for a
massless particle becomes in an equation. For our purposes, the
quantity {\it $A$ must be equal to zero in the massless case,
otherwise, it must be different to zero.}\\
Something between mass and the quantity $A$ is happening, this can be
achieved just saying that $A$ is a function of an small massive test
object $dm$, i.e. a differential of mass:
\begin{eqnarray}
                        A&=&A(dm),
\end{eqnarray}
but, why nature can made a cumbersome function?. Then, let's try 
the easiest one, i.e., a linear function:
\begin{eqnarray}
                        A&=&a_1dm+a_0,
\end{eqnarray}
where $a_1$ and $a_0$ 
are the two constants needed to define a linear function. Using the
ideas described above, we conclude that $a_0=0$. Then let's work out 
the $a_1$ case.\\
Since we are dealing with physics and a constant value, we should take
the fundamental constants of physics in our analyze: 
the Newton's constant $G$, the charge of an electron $e$, 
the speed of light $c$ and the Planck's constant $h$.
From (\ref{a}), using dimensional analysis and the Newton's law of
gravitation, we can argue that $G$, the gravitation constant, is a
good candidate to be the proportional quantity $a_1$. This means that
equation (\ref{a}) turns to:
\begin{eqnarray}
\nonumber       A&=&Gdm\\
                        -dt^2+dx^2+dy^2+dz^2&=&-G^2dm^2
\label{cc}
\end{eqnarray}
The last equation is the equation of an hyper-hyperboloid as it is
shown in figure \ref{lcch} We will call this hyper-hyperboloid as
a {\it causal-hyperboloid}. This is not a hoax, it is just a 
beggarly tentative to approximate causality.  
We can also think that (\ref{cc}) shows another 
dimension associated to the mass of the particle.
\\
Now, we can emphasize the following observations:
\begin{itemize}
\item Absolute past and absolute future are ``connected'' via the
        mass of the particle (see figure \ref{lcch}).
\item Since the value of $G$ is very small, the causal-hyperboloid
        was worked as the light-cone. This case turns back to
        Einstein's Special Relativity.
\item Equation (\ref{cc}) can be written as:
        \begin{equation}
        \label{dS}
             -dt^2+dx^2+dy^2+dz^2+G^2dm^2\doteq d{\cal S}^2,
        \end{equation}
        where we have been introduced an ``new element'' $d{\cal S}^2$.
        In this way, equation (\ref{dS}) defines also a de Sitter space, 
        as will be see in the next lines.
\end{itemize}
To improve a better analyze of (\ref{dS}), we will write it as
$$
  -dt^2+e^{2\beta t}.(dx^2+dy^2+dz^2+G^2dm^2)\doteq d{\cal S}^2,
$$
using
\begin{eqnarray*}
  \beta t&=&ln\beta \frac{v+w}{\alpha},\\
  \beta x&=&\frac{x'}{v+w},\\
  \beta y&=&\frac{y'}{v+w},\\
  \beta z&=&\frac{z'}{v+w},\\
  \beta m&=&\frac{m'}{v+w},
\end{eqnarray*}
the metric goes to
$$
  -dv^2+dw^2+dx'^2+dy'^2+dz'^2+G^2dm'^2=d{\cal S}^2.
$$
It is this last equation we would like to focus on, since it was 
derived from our approach of causality implementation, that we have now
a de Sitter spacetime, see equation (\ref{metric}), but now
in five dimensions ($dS^5$) \cite{dS,large}.
\section{\large A comment on spacetime dimensions.}
\begin{figure}[htb]
\centerline{
\begin{picture}(120,140)(-70,-33)
\Line(-150,10)(150,10)
\put(-75,-10){\makebox(0,0)[br]{$dS^5$}}
\Line(-150,5)(-150,15)
\Line(0,5)(0,15)
\put(75,-10){\makebox(0,0)[br]{$AdS^5$}}
\Line(150,5)(150,15)
\Line(-150,70)(150,70)
\put(-90,85){\makebox(0,0)[br]{$M^4$}}
\Line(-150,65)(-150,75)
\Line(-30,65)(-30,75)
\put(70,85){\makebox(0,0)[br]{${\cal R}^6$}}
\Line(150,65)(150,75)
\put(-140,35){\makebox(0,0)[br]{$D=0$}}
\put(-18,48){\makebox(0,0)[br]{$D=3$}}
\put(12,27){\makebox(0,0)[br]{$D=4$}}
\put(160,35){\makebox(0,0)[br]{$D=9$}}
\end{picture}
}
\caption{Spacetime distribution. $D$ is the dimension and runs from 0 to 9.}
\label{space}
\end{figure}
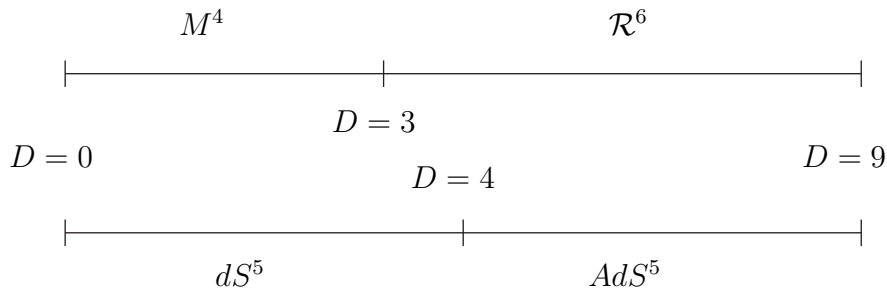

As we have seen in the previous section, the implementation of
causality implies that we have, at least as an approximation, a
de Sitter spacetime instead of the Minkowski space, but now with
five dimensions, i.e., we are dealing with a $dS^5$ space, and this
is a universe of positive cosmological constant $\Lambda$. 
On the other hand an Anti de Sitter space 
in five dimensions, $AdS^5$, describes a 
universe of negative $\Lambda$ and has as interesting
applications in efforts to obtain a better understanding of
string theory \cite{string}. On the way of superstring theory, one
problem consists that the spacetime now has ten dimensions. The idea
is to use the compactification mechanism so finally we have a Minkowski
$M^4$ spacetime and a compactified manifold ${\cal R}^6$ as can be seen
in figure \ref{space}. From the considerations given above, now we have 
a de Sitter spacetime. Then we can fill in the rest of the space 
with another ${\cal R}^5$ space but if we really wish to get something
symmetric, the choice is to pick up the Anti de Sitter spacetime.
working in this manner, we should have a kind of symmetry in spacetime
dimensions, in one hand a de Sitter spacetime with positive $\Lambda$ and
on the other hand a Anti de Sitter spacetime with negative $\Lambda$.
\\
\section{\large Conclusions.}
In this work we just intent to implement causality and observe how this
relates the de Siter spacetime in five dimensions. The idea of more
dimensions is not a newer one but an older \cite{dim} and the pioneers
were Th. Kaluza and O. Klein.
\vskip 1.5cm
{\bf Acknowledgments:} We would like to thanks 
Blanca Zacar\'{\i}as and Daniel Bent\'{\i}n for financial aid.

\end{document}